# A Corrected Parsimony Criterion for Reconstructing Phylogenies

Yue Zhang[1]

Department of Evolution, Ecology, and Organismal Biology, and Museum of Biological Diversity, The Ohio State University, Columbus, OH 43212 USA

## Abstract
In phylogenetic analysis, for non-molecular data, particularly morphology, parsimony optimization is the most commonly employed approach. In the past and present application of the parsimony principle, extra step numbers have been added across different characters without proper justification. This practice, however, has caused the impacts of characters to be inflated or deflated without a valid reason. To resolve this methodological deficiency, here I present a corrected parsimony criterion for reconstructing phylogenies. In essence, character rather than step is the most fundamental unit. Accordingly, the most parsimonious tree should maximize the sum or average of the phylogenetic signals, quantified by retention index, contributed by each character. Assigning proper weights to characters is a separate task that requires information other than the intra-character step number changing range.

## Introduction
The modern field of phylogenetics has been built on the remarkable contribution of Hennig *(1, 2)*, who developed the philosophical framework for reconstructing evolutionary history based on synapomorphies (shared derived traits) among organisms. During the early history of phylogenetics, parsimony analysis of morphological data was the dominant method. For the last few decades, with fast-growing genome sequencing technologies for molecular data, likelihood-based methods have become increasingly common *(3)*. Some researchers have claimed that morphological data are unreliable and can only be used as supplementary to molecular data *(4)*. However, morphology, particularly fossil data from extinct taxa, provides crucial and irreplaceable information for phylogenetic analysis *(5-7)*. Moreover, studies in the field of evolutionary developmental biology have clearly demonstrated that besides genomes, developmental pathways and environmental conditions are also determinant factors driving the evolution of phenotypes *(8-11)*. All information, including molecular, ontogenetic, morphological, behavioral, should be integrated when reconstructing the phylogeny of organisms *(12, 13)*. For non-molecular data, the parsimony method remains the most popular tool. Nonetheless, fundamental flaws have existed in the past-to-present employment of the parsimony method in phylogenetic analysis. Here I discuss the problems in the currently mistakenly-defined "parsimony" principle in phylogenetics, from both theoretical and practical perspectives, using hypothetical and empirical examples. I also present a corrected parsimony criterion and set of procedures that circumvents or minimizes these problems.

[1] Present address: Department of Biology, University of Washington, Seattle, WA 98195 USA
Email: yzfossil@uw.edu

## Logic: characters, character states, steps, and parsimony

At first, it is necessary to clearly understand the connections among characters, character states, and steps. There had been ambiguity, for decades, on the usage of characters and character states in phylogenetics *(14, 15)*. Sereno *(15)* summarized the historical views and presented clear definitions of characters and character states. In essence, characters are heritable organismal features that can be expressed as independent variables. Within each character, mutually exclusive conditions are separated as different character states *(15)*.

In phylogenetic analysis, each character, as an independent variable, provides evidence in discovering the genealogical relationships among the studied taxa. In other words, when each character is viewed alone, the similarity that comes from shared derived character states among different taxa is assumed to be homologous *(16, 17)*. When multiple characters are analyzed simultaneously, conflicts are inevitable. An optimization criterion has to be established to decide which variables are more (or less) reliable for providing hypotheses of homology. When the simplicity rule, a favored philosophical argument across a variety of disciplines, is applied here, it is called the parsimony approach. Apparently, to discover the most parsimonious tree is to find the tree that maximizes homology or minimizes homoplasy.

The key to this issue is how to quantify the amount of homology, or homoplasy, which is measured using extra steps (the number of steps beyond the minimum steps required). Directly counting the number of extra steps across all characters has been employed to quantify the amount of homoplasy for more than half a century since the earliest most "parsimonious" tree *(18)*. A problem with this approach arises because characters differ in their ranges of extra steps. Without recognizing the hierarchical structure of the extra step numbers (intra-character vs. inter-character), when all of the extra steps are added together, the characters with a higher variation range of extra step numbers will cause stronger impacts to the phylogeny compared to those with a lower variation range of extra step numbers.

Here I argue that because each character as an independent variable contributes an independent piece of evidence in reconstructing phylogenies, characters are the minimum indivisible units that should be scored and compared during phylogenetic analysis. When ad hoc hypotheses are minimized, the hierarchical structure of extra steps has to be recognized. Borrowing a core concept from the field of economics, the marginal cost per extra step varies among different characters. The extra step numbers across different characters, therefore, cannot be compared unless they are adjusted to the same scale. The distortion coefficient, *d (19)*, has such a scaling function. $d = \frac{h}{g - m}$, where *g* represents the maximum possible number of steps, *m* represents the minimum possible number of steps, and h represents the number of extra steps *(19)*. The complement of *d*, 1 − *d*, is the retention index *r (20)*, and $r = \frac{g-s}{g-m}$, where *s* represents the number of steps *(20)*. The step count variation range, also the maximum limit of extra step numbers, *g – m*, was referred to as the homoplasy potential *(20)*, here abbreviated as Hp. The most optimal tree minimizes the total amount of homoplasy, ∑*d,* or in other term, maximizes the total expression of the phylogenetic signals, ∑*r*.

Across all characters in any data matrix, *r* ranges from 0 to 1. For the property of *r*, as interpreted by Farris *(20)*, when *r* =1, *s* reaches its minimum limit, *m*. In such a case, "there is no homoplasy" *(20: 417)*. This is the most optimal situation for a character, the phylogenetic signal of which is fully expressed. When *r* = 0, "the character shows as much homoplasy as possible" *(20: 418)*. This is the worst case for a character. "Similarity in this character is then irrelevant to the groupings of the tree" *(20: 418)*. For any value between 0 and 1, "[t]he retention index is then the fraction of apparent synapomorphy in the character that is retained as synapomorphy on the tree" *(20: 418)*. Therefore, for a given character, the expression of its phylogenetic signal can be quantified by the value of *r*.

## Character vs. step in hypothetical examples

The comparison between the parsimony criterion presented here and that used in the current "parsimony" phylogenetics, using hypothetical examples, is presented below, first with a dataset containing binary characters only and then with a dataset containing both binary and multistate characters. These examples will demonstrate the deficiency of the widely used yet mistakenly defined "parsimony" method in phylogenetic analysis.

### Example I: data matrix containing binary characters only

For the data matrix given below, there are six terminal taxa and four binary characters, with Taxon a being the hypothetical ancestor, and state 0 being the more primitive condition. As shown in the matrix, the distribution of the varied states (1 vs. 0) is more skewed in Characters 1 and 4 compared to that in Characters 2 and 3.

|      | a | b | c | d | e | f |
|------|---|---|---|---|---|---|
| Cha1 | 0 | 1 | 1 | 0 | 0 | 0 |
| Cha2 | 0 | 1 | 0 | 1 | 1 | 0 |
| Cha3 | 0 | 0 | 1 | 1 | 0 | 1 |
| Cha4 | 0 | 0 | 0 | 0 | 1 | 1 |

Because of the varied character state distribution patterns, the maximum limit of extra step numbers (Hp) is not uniform across all characters. As a result, among different characters, the impact per extra step to the expression of the phylogenetic signal, quantified by 1/Hp and explained below, is unequal (Table 1).

**Table 1. Comparisons of *m*, *g*, and 1/Hp among each character**

|      | *m* | *g* | 1/Hp |
|------|-----|-----|------|
| Cha1 | 1   | 2   | 1    |
| Cha2 | 1   | 3   | 1/2  |
| Cha3 | 1   | 3   | 1/2  |
| Cha4 | 1   | 2   | 1    |

Using the traditional "parsimony" approach, when all steps are weighted equally, the analysis generates four best trees (Fig. 1, A-D), marked as Trees 1-4. Using the approach present here, when all characters are treated equally, the analysis generates a single best tree (Fig. 1E), marked as Tree*.

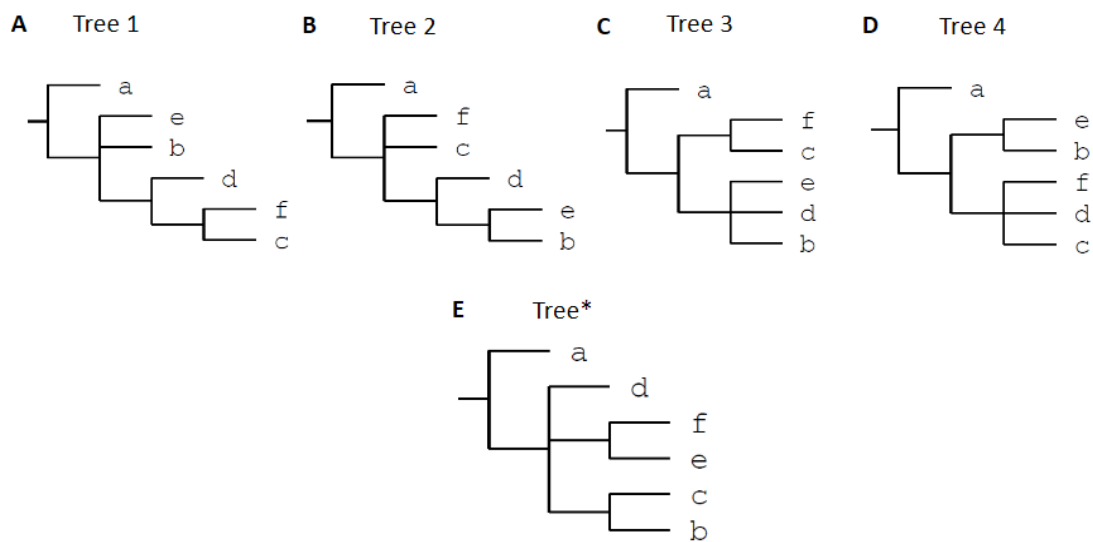

**Fig. 1. The most optimal trees resulted from different parsimony criteria.** (**A-D**) Four best trees (with zero-length branches collapsed) resulted from the traditional "parsimony" approach. (**E**) The single best tree (with zero-length branches collapsed) resulted from the approach that maximizes Σ*r*.

When all characters are treated (weighted) equally, in Tree* (Fig. 1E) Taxa b and c form a clade, as well as Taxa e and f, but neither of these two clades appears in any of the Trees 1-4 (Fig. 1, A-D). Which tree(s) represents the most optimal solution? When extra steps are directly summed without considering the context of characters, Tree* has one step longer in total, which results from two steps shorter in Characters 1 and 4 (one step each) and three steps longer in Characters 2 and 3 (Table 2). Despite the higher total step numbers in Tree*, the impact per extra step to Character 1 or 4 is twice as much as that to Character 2 or 3 (Table 1). One extra step in Character 1 or 4 is enough to turn off the whole phylogenetic signal contributed by the character, whereas one extra step in Character 2 or 3 can only turn off half of its phylogenetic signal (Table 3). Following that each character contributes an independent line of evidence to the phylogeny, it is desirable to make the supported proportion of phylogenetic signal, per character on average ($\bar{r}$), as high as possible. Tree*, therefore, should be preferred compared to Trees 1 to 4 ($\bar{r}$: 1/2 vs. 3/8) (Table 4).

**Table 2. Comparisons of the extra step numbers for each character among different trees.**

|       | Cha1 | Cha2 | Cha3 | Cha4 | Sum |
|-------|------|------|------|------|-----|
| Tree1 | 1    | 1    | 0    | 1    | 3   |
| Tree2 | 1    | 0    | 1    | 1    | 3   |
| Tree3 | 1    | 0    | 1    | 1    | 3   |
| Tree4 | 1    | 1    | 0    | 1    | 3   |
| Tree* | 0    | 2    | 2    | 0    | 4   |

**Table 3. Comparisons of the phylogenetic signal for each character among different trees**

|       | Hypothesis              | Tree 1 | Tree 2 | Tree 3 | Tree 4 | Tree * |
|-------|-------------------------|--------|--------|--------|--------|--------|
| Cha 1 | b & c forming a clade   | X      | X      | X      | X      | O      |
| Cha 2 | b, d, & e forming a clade | O/X  | O      | O      | O/X    | X      |
| Cha 3 | c, d, & f forming a clade | O    | O/X    | O/X    | O      | X      |
| Cha 4 | e & f forming a clade   | X      | X      | X      | X      | O      |

O represents the expression of the hypothesis in the final phylogeny, X represents the non-expression of the hypothesis, and O/X represents that the hypothesis is half expressed.

**Table 4. Comparisons of the *r* values for each character among different trees.**

|       | *r1* | *r2* | *r3* | *r4* | $\sum r$ | $\bar{r}$ |
|-------|------|------|------|------|------|-----|
| Tree1 | 0    | 1/2  | 1    | 0    | 1.5  | 3/8 |
| Tree2 | 0    | 1    | 1/2  | 0    | 1.5  | 3/8 |
| Tree3 | 0    | 1    | 1/2  | 0    | 1.5  | 3/8 |
| Tree4 | 0    | 1/2  | 1    | 0    | 1.5  | 3/8 |
| Tree* | 1    | 0    | 0    | 1    | 2    | 1/2 |

## Example II. A dataset mixed with binary and multistate characters

Below I present a hypothetical data matrix that contains both binary and multistate characters. Characters 1-4 are unordered, and Characters 5-8 are ordered. Taxon a is the outgroup to root the tree.

|      | a | b | c | d | e | f | g | h |
|------|---|---|---|---|---|---|---|---|
| Cha1 | 0 | 0 | 0 | 0 | 0 | 0 | 1 | 1 |
| Cha2 | 0 | 0 | 0 | 0 | 0 | 1 | 1 | 1 |
| Cha3 | 0 | 0 | 0 | 0 | 1 | 1 | 1 | 1 |
| Cha4 | 0 | 0 | 1 | 1 | 2 | 2 | 3 | 3 |
| Cha5 | 0 | 0 | 1 | 1 | 2 | 2 | 3 | 3 |
| Cha6 | 0 | 1 | 0 | 1 | 2 | 3 | 2 | 3 |
| Cha7 | 0 | 1 | 4 | 7 | 6 | 9 | 5 | 1 |
| Cha8 | 0 | 2 | 5 | 8 | 6 | 7 | 9 | 2 |

Table 5 shows the values of $m$, $g$, and Hp for each character. In general, the values of Hp of a character is influenced by the number of character states, the ordered/unordered status, and the character state distribution pattern.

**Table 5. The values of $m$, $g$, and Hp ($g - m$) for each character**

|      | $m$ | $g$ | $g - m$ |
|------|-----|-----|---------|
| Cha1 | 1   | 2   | 1       |
| Cha2 | 1   | 3   | 2       |
| Cha3 | 1   | 4   | 3       |
| Cha4 | 3   | 6   | 3       |
| Cha5 | 3   | 8   | 5       |
| Cha6 | 3   | 8   | 5       |
| Cha7 | 9   | 21  | 12      |
| Cha8 | 9   | 21  | 12      |

For convenience purpose, in the demonstrations here and below, the traditional "parsimony" approach is referred to as flawed parsimony method (FP), and the approach presented in this paper is referred to as the corrected parsimony method (CP). Fig. 2 shows the results of the FP and CP analyses, with each generating a single best tree. Not surprisingly, different optimization criteria generate different best trees. The optimization process, essentially, is for solving the conflicts among different data (characters). To further compare how each character is optimized, extra step numbers and the *r* values of each character are contrasted between the FP and CP trees (Table 6).

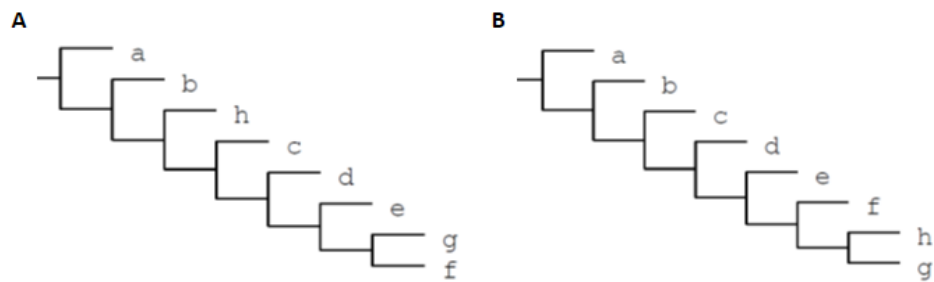

**Fig 2. The single best tree resulted from the FP analysis (A) and the CP analysis (B).**

**Table 6. Comparisons of extra step numbers and *r* values between the FP and CP trees.**

| Cha | *es*(FP) | *es*(CP) | *es*(F − C) | *r*(FP) | *r*(CP) | *r*(F − C) |
|---|---|---|---|---|---|---|
| 1 | 1 | 0 | 1 | 0 | 1 | -1 |
| 2 | 1 | 0 | 1 | 1/2 | 1 | - 1/2 |
| 3 | 1 | 0 | 1 | 2/3 | 1 | - 1/3 |
| 4 | 1 | 0 | 1 | 2/3 | 1 | - 1/3 |
| 5 | 2 | 0 | 2 | 3/5 | 1 | - 2/5 |
| 6 | 3 | 2 | 1 | 2/5 | 3/5 | - 1/5 |
| 7 | 2 | 6 | -4 | 5/6 | 1/2 | 1/3 |
| 8 | 2 | 7 | -5 | 5/6 | 5/12 | 5/12 |
| avg | | | -1/4 | | | -1/4 |

*es*(F − C) denotes the extra step number in the FP tree minus that in the CP tree, for each character. *r*(F − C) denotes that difference for the value of *r*. avg represents the average value.

Compared to the CP tree, on average per character, both of the extra step numbers (*es*) and *r* values are 1/4 lower on the FP tree. Among different characters in a given data matrix, the variation range of extra steps, Hp, theoretically can vary from 0 to infinity. In this case, Hp varies from 1 to 12 (Table 5). Directly comparing the extra step numbers among different characters, the comparison of an object expressed in different scales, is not logically valid. Accordingly, the average 1/4 step shorter in the FP tree bears no meaning. In contrast, the average 1/4 lower on the *r* values shows that on average per character, a quarter of its phylogenetic signal is lost in the FP tree compared to that in the CP tree. Fig. 3 further illustrates the point: one extra step has varied impacts to different characters. For example, comparing the FP tree to the CP tree, one step longer in Character 1 denotes the loss of its full phylogenetic signal, whereas five step shorter in Character 8 only means the gain of less than a half (5/12) of its phylogenetic signal. This hypothetical example, in addition to many previous empirical studies, again shows that under the mistakenly recognized "parsimony" approach, the phylogenetic signal of the whole dataset is often dominated by a few characters with high values of Hp.

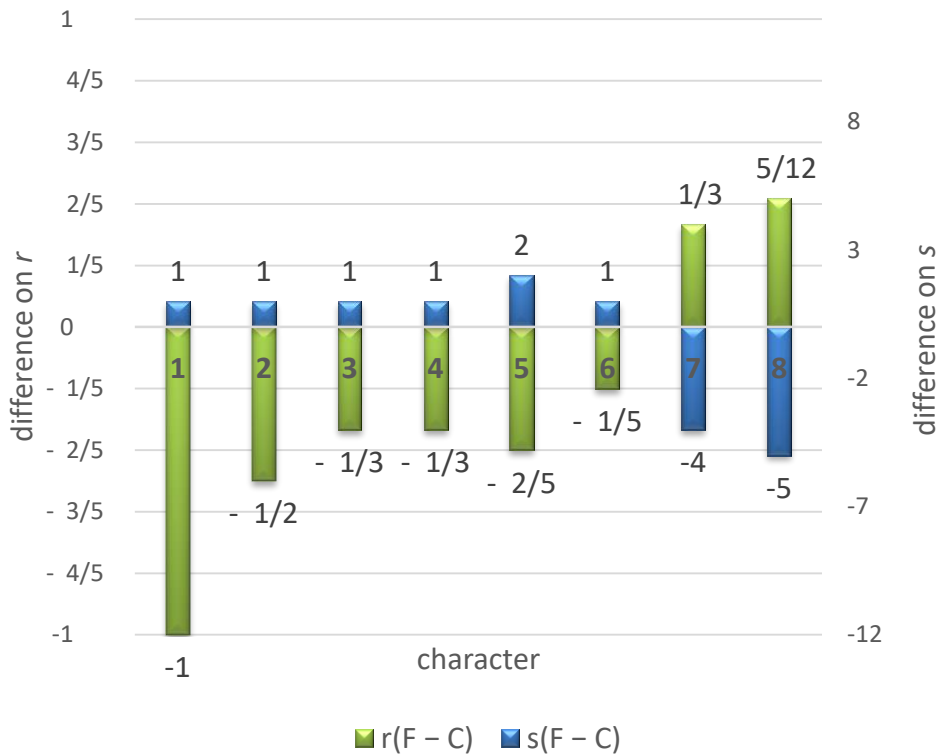

**Fig. 3. Comparisons of the *r* and *s* differences between the FP and CP trees among each character.**

Sereno *(15)* suggested that all morphological characters can be fundamentally classified into two categories: neomorphic and transformational. Apparently, all neomorphic characters, whose states are absent or present, are binary. Transformational characters, depending on the situation, can be either binary or multistate. For the hypothetical example above, without loss of generality, Character 5 is postulated to be an ordered multistate transformational character as described below:

*Feature X, length (L): L<1 cm (0); 1 cm<L<2 cm (1); 2 cm<L<3 mm (2); L>3 cm (3).*

Character 5 is further broken down into three characters below:

*Character 5a: Feature X, length (L) relative to 1 cm: shorter (0); longer (1).*
*Character 5b: Feature X, length (L) relative to 2 cm: shorter (0); longer (1).*
*Character 5c: Feature X, length (L) relative to 3 cm: shorter (0); longer (1).*

Under the currently employed "parsimony" criterion, this conversion generated identical results, not only the identical most "parsimonious" tree, but also the identical tree length. As shown in Table 5, the Hp value is 5 for Characters 5. For Characters 5a, 5b, and 5c, the Hp values are 1, 3, and 1 respectively. For a topology that is given, the extra step cost in Character 5 is identical to that summed from Characters 5a, 5b, and 5c (this result, left for readers to verify, is not listed tree by tree here). In other words, the conversion from Character 5 to Characters 5a, 5b, and 5c, which clearly violates the independency assumption for characters, will cause no effect to the results under the currently accepted "parsimony" criterion. It furthermore demonstrates the fundamental flaw of mixing the intra-character extra step cost with the inter-character extra step cost.

## Character state delimitation issues

Following the hypothetical example incorporating multistate characters, here I further discuss the well-known character state delimitation issue. There are numerous ways to delimit character states for a morphological character, especially "quantitative" characters *(21-24)*. This is one of the most influential issues in phylogenetic analysis using morphological data. Morphological characters are commonly categorized based on qualitative versus quantitative, and discrete versus continuous. These expressions, however, do not represent essential differences among characters *(21)*. It has repeatedly been pointed out that discarding quantitative and continuous data is ungrounded *(15, 21, 24)*. Because most morphological characters describe variations in quantitative traits *(24)*, to minimize information loss, finely-grained character state delimitation approaches are desirable *(21, 24, 25)*.

On the other side, under the current practice of parsimony analysis, finely delimiting character states gives a higher character state number, and thus often a higher impact, to the character *(24, 25)*. As a result, researchers have to confront unlimited ad hoc choices comparing the trade-offs between overweighting characters and losing information *(26)*. Even worse, the result can be easily manipulated by altering the character state delimitation patterns for a few characters.
Using the approach presented here, however, characters' weights will not be amplified or weakened via changing the character state delimitation scheme. Consequently, although it is likely that qualitative characters will "continue to claim a large part of the character data" *(15)*, whenever possible, a finely-grained quantitative character state delimitation should be preferred.

It should be noted that the approach I present here is fundamentally different from those "range-coding" and similar methods proposed by previous researchers *(21, 24, 27)*. Admittedly, those "range-code" *(27)*, "gap-weighting" *(21)*, or similar approaches *(24)* aimed to balance the differential impacts among characters by downweighting multistate characters. There is, however, a common flaw for those methods. Because the variation range of extra step numbers (Hp), rather than the states count, is the determinant factor to the influence of a character, "a binary character with low consistency may well have more steps than a multistate character" *(28: 95)*.

# Further discussion on weighting characters

To sum up the rationale of the corrected parsimony criterion presented here, characters rather than steps are the minimum independent functional units in phylogenetic analysis. Therefore, characters are the raw materials that determine steps, not the other way around. The causal relationship between characters and the changing ranges of extra steps determines the hierarchical structure of extra steps, intra-character vs. inter-character, during phylogenetic analysis. It is necessary to understand that characters are the core contents in phylogenetic analysis, and extra steps are the medium. Extra step numbers are counted for calculating how well the phylogenetic signal of a character is supported by others. Leaving out the context of characters, merely counting extra step numbers can hardly be meaningful.

How each character should be weighted is another question. Theoretically, because each character represents an indivisible independent variable, it makes sense to unweight, or uniweight, each character, as shown in the hypothetical examples above. In practice, however, "there seems to be no sensible reason to insist that all characters should make the same contribution" *(28: 95)*. As previous researchers have pointed out *(15, 22, 29, 30, 31)*, there are many factors (such as missing data, polymorphism, and character correlation) that can potentially influence the reliability of a character. Therefore, it is especially important to make decisions *a priori*, during the character analysis stage, with explicitly presented reasons *(22, 29, 30, 31)*. How much qualitative/quantitative weight should be given to, including whether retaining (weight > 0) or discarding (weight = 0), a character? It remains one of the most challenging issues in phylogenetic analysis. I certainly do not agree that all characters deserve the same weight. Rather, characters should be weighed differentially, as long as the weights are supported by legitimate reasons. The point is: given that Character A has a higher range of extra step numbers than Character B, it is not a legitimate reason for Character A to be assigned a higher weight than Character B.

# Mathematical expression

As shown here, unweighting characters demands weighting steps, but unweighting steps results in weighting characters. Above, the relationship between characters and steps is interpreted on a logical basis, accompanied by hypothetical examples; below, it is denoted in mathematical forms.

For a data matrix that includes n characters:

$$\sum_{i=1}^{n} r_i = \sum_{i=1}^{n} \frac{g_i - s_i}{g_i - m_i} = \sum_{i=1}^{n} \frac{g_i}{g_i - m_i} - \sum_{i=1}^{n} \frac{s_i}{g_i - m_i} \quad \ldots\ldots\ldots(1)$$

where $r_i$, $g_i$, $m_i$, and $s_i$ are the retention index, the maximum possible step numbers, the minimum possible step numbers, and the step numbers for character *i*, respectively.

For a given data matrix, the value of $\sum_{i=1}^{n} \frac{g_i}{g_i - m_i}$ is a constant. Hence, maximizing $\sum_{i=1}^{n} r_i$ equals minimizing $\sum_{i=1}^{n} \frac{s_i}{g_i - m_i}$, which is interpreted as minimizing the step numbers, with each step weighted by 1/Hp. On the contrary, the mistakenly defined "parsimony" analysis searches for the tree that minimizes the direct sum of step numbers, $\sum_{i=1}^{n} s_i$.

$$\sum_{i=1}^{n} s_i = \sum_{i=1}^{n}[(g_i - m_i) \cdot \frac{s_i}{g_i - m_i}] = \sum_{i=1}^{n}[(g_i - m_i) \cdot (\frac{g_i}{g_i - m_i} - \frac{g_i - s_i}{g_i - m_i})]$$

$$= \sum_{i=1}^{n}[(g_i - m_i) \cdot (\frac{g_i}{g_i - m_i}) - (g_i - m_i) \cdot \frac{g_i - s_i}{g_i - m_i})]$$

$$= \sum_{i=1}^{n} g_i - \sum_{i=1}^{n} (g_i - m_i) \cdot r_i \quad \ldots\ldots\ldots\ldots\ldots\ldots\ldots\ldots\ldots\ldots\ldots\ldots\ldots\ldots\ldots\ldots\ldots\ldots(2)$$

For a given data matrix, the value of $\sum_{i=1}^{n} g_i$ is a constant. Hence, minimizing $\sum_{i=1}^{n} s_i$ equals maximizing $\sum_{i=1}^{n} (g_i - m_i) \cdot r_i$, which can be interpreted as maximizing the total expression of the phylogenetic signals, with each character weighted by Hp.

The calculation of $\sum r$ or $\bar{r}$ has been employed for comparing the homoplasy level among different sets of characters on a given tree *(32, 33)*, or among a set of mistakenly called most "parsimonious" trees *(34)*, but has not been used for the purpose stated here. The approach of maximizing $\sum r$, however, has been criticized for giving "higher 'weight' to characters with less informative variation" *(35: 86)*. The "informative variation" *(35, 36)* was referred to as $g - m$, which is Hp, and the "weight" *(35)* was referred to as $\frac{1}{g - m}$. It is, however, the weight to steps, not to characters. Maximizing $\sum r$ does not give higher or lower weight to characters with less or more Hp. It gives equal weight to each character, which requires steps be weighted by 1/Hp. Conversely, the traditional "parsimony" approach, by indiscriminately counting steps across different characters, gives higher weight to characters with higher values of Hp. The "informative variation" from one character, before recognized as homology or homoplasy, is subjected to test among those from others. A character with higher "informative variation" does not warrant that the information from this character is more reliable without testing.

# Demonstration using empirical case

To further show the fallacy in the traditional mistakenly defined "parsimony" method, a published empirical phylogenetic dataset of hominoids from Strait and Grine (SG) *(37)* is examined here. The analysis by SG included two outgroup taxa (i.e., *Colobus* and *Papio*), five extant hominoid taxa (i.e., *Hylobates*, *Pongo*, *Gorilla*, *Pan*, and *Homo sapiens*), and 13 extinct hominid taxa *(37)*. In the demonstration below, for simplification purpose, only the extant taxa were included. The source dataset included 198 characters, of which 109 were "traditional characters" and 89 were "craniometric characters" *(37)*. As SG described: "Traditional characters may be quantitative or qualitative, but when they are quantitative, they typically measure an aspect of morphology that has also been qualitatively described" *(37: 405)*, whereas "[c]raniometric characters represent size-adjusted linear dimensions measured between standard cranial landmarks" *(37: 405)*.

Character states were delimited differentially between the two categories. "[S]tates for all of the traditional characters examined here (both qualitative and quantitative) were determined in a similar fashion" *(37: 427)*. The "qualitative traditional characters" were delimited by frequency *(38)*, and the "quantitative traditional characters" were delimited using the range-based method *(39)*. For the craniometric characters, the states "were obtained using homogeneous subset coding (HSC) *(40)*" *(37: 427)*. Details of the HSC method *(39)* is not repeated here. In simple term, the HSC method *(40)* is a finer tool to delimit character states compared to the range-based method *(39)*. As a result, the step number changing range (Hp) of "craniometric characters" tend to be larger than that of "traditional characters". Among all of the phylogenetically informative characters in the dataset, the Hp values vary from 1 to 9.

All data (Data A1) is copied directly from the source *(37)* without modification. In particular, in the original analysis by SG, craniometric characters that described different parts of the same character complex were downweighted *(37: 426: Table 6)*. Such adjustment was done for the purpose of preventing redundant calculations of correlated characters and is followed here (Table A1) in the demonstration below in both of the FP and CP analyses.

## Comparison 1: separate groups based on Hp vs. total data

In this comparison, the characters are divided into nine groups based on their Hp values (Table A1). Because steps among different characters will bear the same weight if the characters analyzed have the same value of Hp, when each of the nine character groups is analyzed separately, the two methods compared here (FP and CP) always yield congruent results. The trees from these nine separate analyses are then combined in two different ways, biased and unbiased, as explained below. The consensus trees are subsequently compared to the CP and FP trees from the whole data respectively.

The rationale is provided here. The most parsimonious tree from a given group, as the most optimal combination of all phylogenetic signals from the characters within that group, is used to roughly represent the aggregated phylogenetic signal from the group.

The best tree of the whole dataset can be viewed as a compromised optimum among the phylogenetic signals from different groups. The consensus tree, with the separate trees from each group assigned appropriate weights according to the strength of their phylogenetic signals, is expected to be similar to the best tree yielded from the whole dataset.

The strength of the phylogenetic signal from each group can be denoted by the product of two factors: the average phylogenetic signal expressed per character ($\bar{r}$) and the number of characters. For example, considering two hypothetical character groups, their strengths in interpreting the phylogenetic relationships among the same taxa are contrasted as below. Group 1 has 100 characters and Group 2 has 50 characters. $\bar{r}$ is 0.2 in Group 1, and 0.5 in Group 2. Then, among the 100 characters in Group 1, 80% of their phylogenetic signals were neutralized due to the conflicts among themselves; whereas among the 50 characters in Group 2, 50% of their phylogenetic signals were neutralized. Consequently, the strength ratio of the aggregated phylogenetic signal from Group 1 to that from Group 2 is: $\frac{100 \times 0.2}{50 \times 0.5} = \frac{4}{5}$.

For the empirical case discussed here, using $N_i$ to represent the number of independent characters [after the downweighting adjustment for redundant characters *(37)*] in group $i$, $\bar{r}_i$ to represent the average value of $r$ in group $i$, then, if each independent line of evidence is treated in an unbiased way, the weight (strength ratio) of the phylogenetic signal from Groups 1 to 9 is
$\bar{r}_1 N_1 : \bar{r}_2 N_2 : \bar{r}_3 N_3 : \bar{r}_4 N_4 : \bar{r}_5 N_5 : \bar{r}_6 N_6 : \bar{r}_7 N_7 : \bar{r}_8 N_8 : \bar{r}_9 N_9$. However, if characters are weighted in a biased way, the weight of the phylogenetic signal from each group will be tilted. In the FP method, the ratio of the effective character numbers from Group 1 to 9 are tilted to $N_1 : 2N_2 : 3N_3 : 4N_4 : 5N_5 : 6N_6 : 7N_7 : 8N_8 : 9N_9$, and the weights of the phylogenetic signal from each group are tilted accordingly.

When each character group was analyzed apart, each separate analysis generated only one most parsimonious tree except for the analysis of Group 3, for which the corresponding weight is equally divided between the two equally most parsimonious trees. Table 7 shows the data from each group. The trees resulting from the separate analyses are combined in two different ways to calculate the consensus tree, based on the unbiased and biased weights of the phylogenetic signals from each group (Table 7), as explained above.

**Table 7. Data from each character group**

| Group | Hp | $\bar{r}$ | Number of characters | Biased number of characters | Unbiased weight of phylogenetic signal | Biased weight |
|---|---|---|---|---|---|---|
| 1 | 1 | 0.364542 | 20.9 | 20.9 | 16 | 5 |
| 2 | 2 | 0.476643 | 31.6 | 63.2 | 33 | 18 |
| 3 | 3 | 0.460827 | 13.7 | 41.2 | 14 | 12 |
| 4 | 4 | 0.393869 | 11.6 | 46.2 | 10 | 11 |
| 5 | 5 | 0.308772 | 9.5 | 47.5 | 6 | 9 |
| 6 | 6 | 0.343915 | 5.3 | 31.5 | 4 | 7 |
| 7 | 7 | 0.472947 | 5.8 | 40.4 | 6 | 12 |
| 8 | 8 | 0.557143 | 5.8 | 46.7 | 7 | 16 |
| 9 | 9 | 0.568228 | 3.6 | 32.4 | 4 | 11 |

For the comparison of the similarity between the consensus trees and the trees from the whole data, the normalized Robinson-Foulds Distance *(41)* (RFD) was calculated using the software TNT *(42)*. The RFDs between the CP tree (Fig. 7B) and the consensus trees, with the trees from separate groups combined in the unbiased (Fig. 4A) and biased (Fig. 4C) ways, are 0.2 and 0.4 respectively. Such values for the FP tree (Fig. 4D) are 0.4 and 0.2 respectively. This result reveals that the FP method weights characters in a biased way, as presented in Equation (2) and demonstrated here.

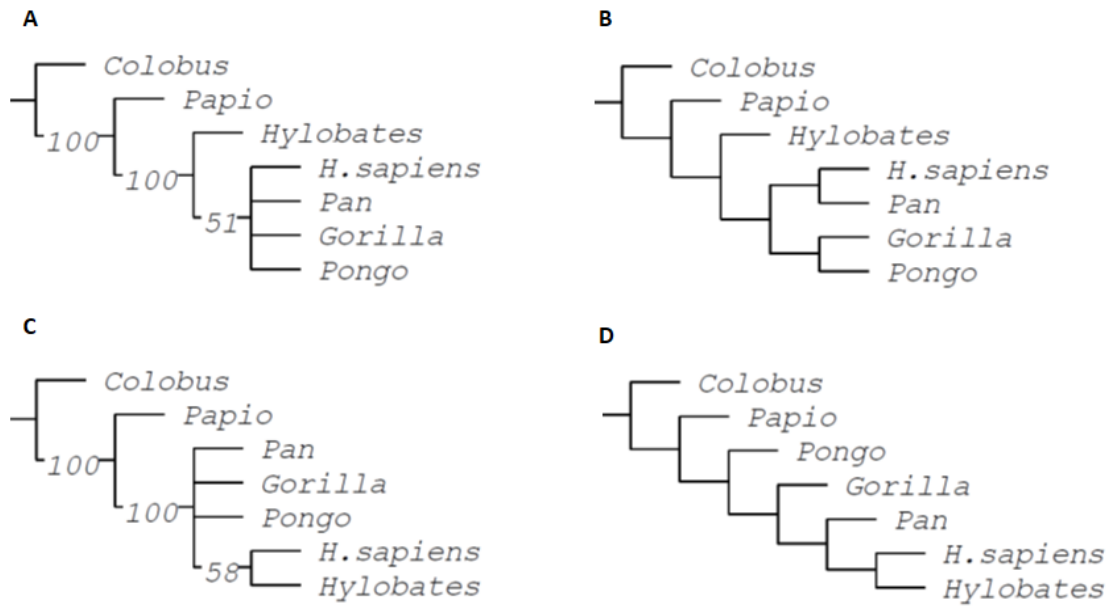

**Fig. 4. The hominoid trees in Comparison 1.** (**A**) The majority rule consensus tree with the best trees from each separate group combined in the unbiased way. (**B**) The most parsimonious tree from the total data using the CP approach. (**C**) The majority rule consensus tree with the best trees from each separate group combined in the biased way. (**D**) The most parsimonious tree from the total data using the FP approach.

## Comparison 2: up-act and down-act vs. total data

The rationale of Comparison 2 is similar to that of Comparison 1, and the nine separate character groups are still used. In Comparison 1, nine separate analyses are conducted with only one character group activated in each analysis. In Comparison 2, a method that I named up-act and down-act is applied, as shown in Table 8. In the process of up-act, there are eight separate analyses. In the first analysis, Up-act 1, Groups 2-9 (upper than 1) are activated, leaving Group 1 deactivated. In the second analysis, Up-act 2, Groups 3-9 (upper than 2) are activated, and so on, until Analysis Up-act 8, when only Group 9 is activated. In the process of down-act, there are also eight separate analyses. Similarly, in the Analysis Down-act n (n = 9, 8…2), groups lower than n are activated, leaving other groups deactivated.

**Table 8. The activated character groups in each separate analysis in the up-act and down-act process**

| Separate analysis | Groups activated |
|---|---|
| Up-act 1 (Ua1) | 2, 3, 4, 5, 6, 7, 8, 9 |
| Up-act 2 (Ua2) | 3, 4, 5, 6, 7, 8, 9 |
| Up-act 3 (Ua3) | 4, 5, 6, 7, 8, 9 |
| Up-act 4 (Ua4) | 5, 6, 7, 8, 9 |
| Up-act 5 (Ua5) | 6, 7, 8, 9 |
| Up-act 6 (Ua6) | 7, 8, 9 |
| Up-act 7 (Ua7) | 8, 9 |
| Up-act 8 (Ua8) | 9 |
| Down-act 9 (Da9) | 1, 2, 3, 4, 5, 6, 7, 8 |
| Down-act 8 (Da8) | 1, 2, 3, 4, 5, 6, 7 |
| Down-act 7 (Da7) | 1, 2, 3, 4, 5, 6 |
| Down-act 6 (Da6) | 1, 2, 3, 4, 5 |
| Down-act 5 (Da5) | 1, 2, 3, 4 |
| Down-act 4 (Da4) | 1, 2, 3 |
| Down-act 3 (Da3) | 1, 2 |
| Down-act 2 (Da2) | 1 |

For each of the 16 separate analyses, the most optimal tree (under the specific optimization criterion) is used to represent the phylogenetic signal from the corresponding dataset. All of these trees are then combined to calculate the consensus tree. When all of the separate datasets are summed, each character group is counted eight times. Hence, the consensus tree, where the redundant counts for each group are counterbalanced by each other, is comparable to the tree from the total data.

To explain the calculation of the phylogenetic signal weights from each group, here I still use $N_i$ to represent the number of characters in group $i$, and $\bar{r}_i$ to represent the average value of $r$ in group $i$. Among the FP analyses, for example, when the dataset from Analysis Ua5 is combined with that from Analysis Da6, the strengths of the phylogenetic signal from the activated groups in Analysis Ua5 (Groups 6-9) is compared with that from the activated groups in Analysis Da6 (Groups 1-5). The ratio of the number of characters that take effect is $\dfrac{6 \cdot N_6 + 7 \cdot N_7 + 8 \cdot N_8 + 9 \cdot N_9}{N_1 + 2 \cdot N_2 + 3 \cdot N_3 + 4 \cdot N_4 + 5 \cdot N_5}$.

Correspondingly, in Analysis Ua5, the average expression of the phylogenetic signal per effective character is $\dfrac{6 \cdot \bar{r}_6 \cdot N_6 + 7 \cdot \bar{r}_7 \cdot N_7 + 8 \cdot \bar{r}_8 \cdot N_8 + 9 \cdot \bar{r}_9 \cdot N_9}{6 \cdot N_6 + 7 \cdot N_7 + 8 \cdot N_8 + 9 \cdot N_9}$. In Analysis Da6, it is $\dfrac{\bar{r}_1 \cdot N_1 + 2 \cdot \bar{r}_2 \cdot N_2 + 3 \cdot \bar{r}_3 \cdot N_3 + 4 \cdot \bar{r}_4 \cdot N_4 + 5 \cdot \bar{r}_5 \cdot N_5}{N_1 + 2 \cdot N_2 + 3 \cdot N_3 + 4 \cdot N_4 + 5 \cdot N_5}$.

Among the CP analyses, because of the equal impact of each character, the calculation of the phylogenetic signal weights from each separate dataset is straightforward. For the same example, the ratio of the number of characters that take effect between Analyses Ua5 and Da6 is $\frac{N_6 + N_7 + N_8 + N_9}{N_1 + N_2 + N_3 + N_4 + N_5}$, and the average expression of the phylogenetic signal per effective character is simply $\bar{r}$.

Accordingly, the trees from the separate CP analyses were combined based on the strength ratio of the phylogenetic signals among each separate dataset under the CP scheme, here referred to as the unbiased way. The trees from the separate FP analyses were combined based on the strength ratio of the phylogenetic signals among each separate dataset under the FP scheme, referred to as the biased way. For comparison purpose, the separate FP trees were also combined in the unbiased way, here referred to as the way of "if-not-biased".

Each separate analysis generated only one best tree. The TNT scripts written for the calculations and the results in detail are presented in the supplementary materials (Scripts A1-A8, see Appendix for details). Tables 9 and 10 show the data from each separate analysis, using the CP and FP approach respectively.

**Table 9. Data from each separate analysis using the CP approach**

| Separate analysis | $\bar{r}$ | Number of characters | Weight of phylogenetic signal |
|---|---|---|---|
| Ua1 | 0.359082 | 86.8 | 10 |
| Ua2 | 0.366492 | 55.2 | 6 |
| Ua3 | 0.381626 | 41.5 | 5 |
| Ua4 | 0.383381 | 30.0 | 4 |
| Ua5 | 0.431075 | 20.4 | 3 |
| Ua6 | 0.524213 | 15.2 | 2 |
| Ua7 | 0.555556 | 9.4 | 2 |
| Ua8 | 0.568228 | 3.6 | 1 |
| Da9 | 0.341303 | 104.1 | 11 |
| Da8 | 0.343966 | 98.3 | 11 |
| Da7 | 0.350288 | 92.5 | 10 |
| Da6 | 0.356081 | 87.3 | 10 |
| Da5 | 0.37514 | 77.8 | 9 |
| Da4 | 0.40035 | 66.2 | 8 |
| Da3 | 0.403434 | 52.5 | 7 |
| Da2 | 0.364542 | 20.9 | 2 |

**Table 10. Data from each separate analysis using the FP approach**

| Separate analysis | "if-not-biased" $\bar{r}$ | "if-not-biased" number of characters | "if-non-biased" weight of phylogenetic signal | Biased $\bar{r}$ | Biased number of characters | Biased weight of phylogenetic signal |
|---|---|---|---|---|---|---|
| Ua1 | 0.34204  | 86.8  | 10 | 0.373037 | 349.0 | 11 |
| Ua2 | 0.366492 | 55.2  | 7  | 0.389384 | 285.8 | 10 |
| Ua3 | 0.381626 | 41.5  | 5  | 0.399402 | 244.7 | 9  |
| Ua4 | 0.383381 | 30.0  | 4  | 0.410348 | 198.5 | 7  |
| Ua5 | 0.431075 | 20.4  | 3  | 0.451144 | 151.0 | 6  |
| Ua6 | 0.524213 | 15.2  | 3  | 0.527556 | 119.5 | 6  |
| Ua7 | 0.555556 | 9.4   | 2  | 0.555438 | 79.1  | 4  |
| Ua8 | 0.568228 | 3.6   | 1  | 0.568228 | 32.4  | 2  |
| Da9 | 0.294857 | 104.1 | 10 | 0.341568 | 337.6 | 10 |
| Da8 | 0.332654 | 98.3  | 11 | 0.328572 | 290.9 | 8  |
| Da7 | 0.34964  | 92.5  | 10 | 0.329326 | 250.5 | 7  |
| Da6 | 0.355393 | 87.3  | 10 | 0.340164 | 219.0 | 7  |
| Da5 | 0.374369 | 77.8  | 9  | 0.378982 | 171.5 | 6  |
| Da4 | 0.395319 | 66.2  | 8  | 0.424988 | 125.3 | 5  |
| Da3 | 0.384393 | 52.5  | 6  | 0.419049 | 84.1  | 3  |
| Da2 | 0.364542 | 20.9  | 2  | 0.364542 | 20.9  | 1  |

The CP tree from the whole data (Fig. 5D) agrees with the majority rule consensus tree that combines the separate CP trees in the unbiased way (Fig. 5A). In contrast, when the separate FP trees are combined in the same unbiased way, the "if-not-biased" way, the majority rule consensus tree (Fig. 5B) differs significantly from the FP tree yielded from the whole data (Fig. 5E). The RFD between these two trees is 0.40. Whereas, when the separate FP trees are combined in the biased way according to the impacts of separate datasets under the FP scheme, the resulting majority rule consensus tree (Fig. 5C) is much closer to the FP tree from the whole data (Fig. 5E), with an RFD of 0.10. Such a result is expected, because the FP scheme does treat characters in a biased way.

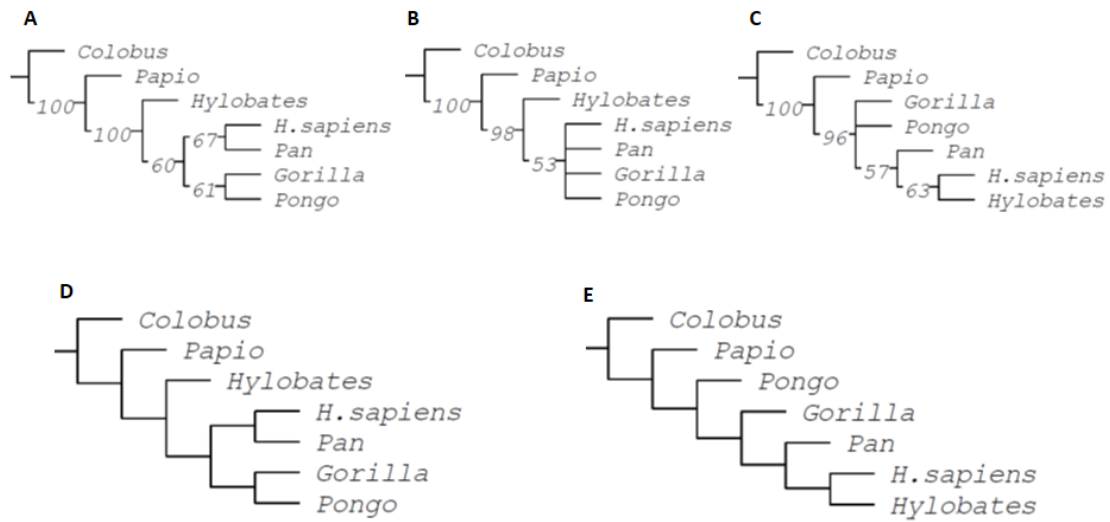

**Fig. 5. The hominoid trees in Comparison 2.** (**A**) The majority rule consensus tree from the 16 separate trees using the CP approach, combined in the unbiased way. (**B**) The majority rule consensus tree from the 16 separate trees using the FP approach, combined in the "if-not-biased" way. (**C**) The majority rule consensus tree from the 16 separate trees using the FP approach, combined in the biased way. (**D**) The most parsimonious tree from the total data using the CP approach. (**E**) The "most parsimonious" tree from the total data using the FP approach.

Comparisons 1 and 2 presented above both demonstrate that the traditional "parsimony" approach weights characters in proportion to their Hp values, as shown in Equation (2). In this empirical example, the heterogeneity of Hp values is mainly attributed to the diverse character state delimitation patterns. Finely delimiting character states is merited and should be employed when the situation permits, as discussed earlier. It is unjustified, however, to claim that characters with finely-grained character states are automatically more important than those with coarsely-grained character states in interpreting the phylogenetic relationships.

The empirical hominoid dataset *(37)* used here, composed of both qualitative and quantitative characters, represents a typical morphological dataset in phylogenetic analysis. Using the case here, as a simple example, the character "Malar prognathism" *(37)* has an Hp value of 9 (Table A1: Character 52), and the character "Relative enamel thickness" *(37)* has an Hp value of 1 (Table A1: Character 147). Under the traditional "parsimony" principle, the impact of the former is nine times as strong as that of the latter. This difference, not reflecting the differential reliability between these two characters, merely comes from the fact that the former character was quantified using morphometric data whereas the latter was described qualitatively (i.e., thin vs. thick).

Supposing that the data are reanalyzed, for the character "Malar prognathism", the data are delimited using a finer approach (e.g., delimited "as such" on a finer measurement scale). It will be easy to make the impact of this character, rather than nine times, perhaps 90 times as strong as that of the character "Relatively enamel thickness". In another scenario, if the character "Relative enamel thickness" is quantified and finely delimited, the situation can be easily reversed. The advantage of quantifying and finely delimiting characters, increasing the resolution, is obvious. Meanwhile, given a character, researchers have to distinguish the difference between the resolution and the reliability of the data.

Does the character "Malar prognathism" deserve nine times weight as the character "Relatively enamel thickness"? The answer is, depending on additional information, maybe or maybe not. Supposing a scenario that the former character has been well studied across all the taxa in the analysis whereas the latter character has only been studied on few samples from a fraction of the taxa, then it should be claimed that the former character deserves a higher weight than the latter (not necessarily 9:1). Such claim, however, would be based on the differential levels of study on these two characters, rather than that one is quantified whereas the other is not. Despite that quantification may imply a better study in some cases, it is not necessarily so. It is certainly possible that the data of a qualitatively described character is derived from a more reliable study compared to that of a quantitatively described character, in which case a qualitatively described character may deserve a higher weight.

Apparently, a thorough character analysis, including differentially weighting characters *a priori* based on legitimate reasons, is desirable and necessary. Many factors can affect a character's weight *a priori*, and it should be discussed case by case *(15, 22, 29, 30, 31)*. When such specific information is not analyzed, solely from the theoretical basis of parsimony principle, indiscriminately treating each character as each independent line of evidence should be more rational compared to differentially weighting characters based on unjustifiable reasons (e.g., the different Hp values). Another thing to note here is that in the empirical example above, although we may never be sure what the true phylogeny is, we should be fairly confident that we humans are more closely related to chimpanzees, as shown in the CP tree, than to gibbons, as shown in the FP tree. This point is not discussed here because it has little to do with which of the approaches, CP vs. FP, is better. Because in phylogenetic analysis the data is always incomplete and never error-free, it is certainly possible to obtain a good tree (or even the true tree if we may know) using a flawed approach. Regardless of whatever result a method could generate in whatever case, if the method shows flaws in its logic, it has to be corrected.

## Conclusion

Reconstructing phylogeny is essential in evolutionary biology. Among the different approaches, parsimony remains one of the most powerful tools, particularly for non-molecular data. The currently employed "parsimony" method, lasting for more than half a century, however, is built on a flawed understanding of the criterion. This paper presents the causal relationship between characters and extra steps. Accordingly, character, as the core content in phylogenetic reconstruction, should be treated as the minimum independent element. When the number of extra steps is calculated, the relative rather than the absolute values should be used. This correction in the principle of reconstructing the most parsimonious tree is necessary, in that it prevents a character's impact from being magnified or depressed solely based on a higher or lower value of Hp. Characters' reliability, one from another, is expected not to be all the same. The difference, however, is attributed to factors other than the value of Hp, such as character correlation, data incompleteness, sample limitation, and polymorphism. How to weight characters, with well-argued reasons, remains one of the most challenging tasks in phylogenetics. To accurately weight different characters is expected to be a process that involves continuous "checking, correcting, and rechecking" *(2)*. The corrected parsimony criterion, mainly clarified from the theoretical basis in contrast to the mistakenly defined "parsimony" method, is presented here as a separate issue from character weighting. Apparently, when character weights can be clearly assigned with grounded reasons, the calculation of summing or averaging the retention indices should take into account of the specified character weights. Given a dataset for parsimony phylogenetic analysis, the average retention index, in addition to being used for discovering the most parsimonious tree, can also reflect the level of compatibility among all characters within the same dataset. Furthermore, the average retention index can be compared across different, even nonrelated, datasets in interpreting which dataset is more self-compatible versus another.

The parsimony criterion, treating each character as an indivisible unit and maximizing the summed phylogenetic signal contributed by each character, is mainly discussed here based on morphological data. The basic principle should be applicable to all types of data. Additionally, when the parsimony phylogenetic approach is used beyond the field of biology, such as archaeology, psychology, and sociology, it is expected that the general logic can be similarly applied.

## Acknowledgments

This manuscript derives from a chapter of my dissertation project that was accomplished in 2015. I thank the comments, discussions, and critiques from my former advisor John Hunter and my former dissertation committee: John Freudenstein, Bill Ausich, and Debbie Guatelli-Sternberg. I also thank Jin Meng, John Wenzel, Greg Wilson, Tony Harper, Chris Brochu, Charles Marshall, and anonymous reviewers. In particular, I am deeply indebted to Professor Richard C. Fox, for the invaluable critiques, encouragements, and all the help I have benefited from him.

# Appendix

**Part I. TNT Scripts and notes**

All of the scripts were prepared for running parsimony analysis using the program TNT v1.1 *(42)*. The scripts are divided into two main categories. In the empirical example of the hominoid phylogeny, following the original study by SG *(37)*, some of the craniometric characters were downweighted due to redundancy *(37: 426: Table 6)*. The scripts for analyzing the SG dataset in particular are thus different from the scripts for analyzing a dataset in general, in which no weighting adjustment is assumed to be made *a priori*. For the TNT scripts presented below, Scripts A1-A2 are for analyzing a data matrix in a general case, and Scripts A3-A8 are for analyzing the specific SG data matrix *(37)* (Data A1), as discussed in the main text. All TNT scripts (.run) below are presented in plain text format.

## 1.1. Scripts for general data matrices

**Script A1. CP.run:**
The script for weighing steps in proportion to 1/Hp, as the parsimony criterion presented in the main text.

```
macro= ;
loop 0 nchar
    if (( maxsteps[#1] - minsteps[#1] )==0)
     ccode ] #1;
    else set 1  1 / ( maxsteps[#1] - minsteps[#1] );
macfloat 3;
set 1 '1' * 1000 ;
ccode /'1' #1 ;
end
stop;
end ;
```

**Script A2. AR.run:**
The scripts for reporting the average *r* value among all characters, on the last tree in memory if multiple trees are saved from previous searches.

```
macro= ;
macfloat 6;
    set 3 0;
    set 4 0;
loop 0 nchar
    if (!isact[#1]) continue; end
    set 1 length[0 #1]/weight[#1];
    set 2 ( maxsteps[#1] - '1' ) / ( maxsteps[#1] - minsteps[#1] );
    set 3 '3'+1;
    set 4 '4'+'2';
    stop
   set 7 '4'/'3';
   quote the averaged ri is '7' ;
proc/;
```

## 1.2. Scripts for the empirical example using the SG matrix

**Script A3. CPSG.run:**
The script for weighting steps under the parsimony criterion presented in the main text, for the SG matrix specifically.

```
macro= ;
 loop 0 nchar
    if (( maxsteps[#1] - minsteps[#1] )==0)
     ccode ] #1;
    else set 1 weight[#1] ;
     macfloat 3;
     set 2  1 / ( maxsteps[#1] - minsteps[#1] );
      set 3 '1' * '2' ;
  macfloat 0;
ccode /'3' #1 ;
end
stop;
end ;
```

**Script A4. ARSG.run:**
The script for reporting the average *r* value for the SG dataset phylogenies, on the last tree in memory.

```
macro= ;
macfloat 6;
    set 3 0;
    set 4 0;
loop 0 nchar
    if (!isact[#1]) continue; end
    set 1 length[0 #1]/weight[#1];
    set 2 (maxsteps[#1] - '1')/(maxsteps[#1] - minsteps[#1]);
     if
((#1==0)||(#1==1)||(#1==2)||(#1==3)||(#1==28)||(#1==138)||(#1==7)||(
#1==8)||(#1==9)||(#1==10)||(#1==29)||(#1==141))
     set 5 1/6;
     set 6 '2'*'5';
     set 3 '3'+'5';
     set 4 '4'+'6';
     continue;
     end
     if ((#1==4)||(#1==5)||(#1==6)||(#1==139))
     set 5 1/4;
     set 6 '2'*'5';
     set 3 '3'+'5';
     set 4 '4'+'6';
     continue;
      end
       if
((#1==11)||(#1==12)||(#1==13)||(#1==14)||(#1==15)||(#1==16)||(#1==30
)||(#1==142))
     set 5 1/8;
    set 6 '2'*'5';
     set 3 '3'+'5';
     set 4 '4'+'6';
     continue;
```

```
        end
        if
((#1==17)||(#1==18)||(#1==20)||(#1==31)||(#1==34)||(#1==41)||(#1==35)||(#1==36)||(#1==89)||(#1==39)||(#1==40)||(#1==117)||(#1==56)||(#1==61)||(#1==114)||(#1==67)||(#1==68)||(#1==69))
        set 5 1/3;
        set 6 '2'*'5';
        set 3 '3'+'5';
        set 4 '4'+'6';
        continue;
        end
        if
((#1==19)||(#1==22)||(#1==21)||(#1==26)||(#1==37)||(#1==43)||(#1==48)||(#1==49)||(#1==85)||(#1==87)||(#1==65)||(#1==66)||(#1==75)||(#1==76)||(#1==77)||(#1==78))
        set 5 1/2;
      set 6 '2'*'5';
        set 3 '3'+'5';
        set 4 '4'+'6';
        continue;
          end
          if
((#1==55)||(#1==64)||(#1==79)||(#1==82)||(#1==83)||(#1==84)||(#1==86))
        set 5 1/7;
        set 6 '2'*'5';
        set 3 '3'+'5';
        set 4 '4'+'6';
        continue;
        end
      set 3 '3'+1;
      set 4 '4'+'2';
      stop

    set 7 '4'/'3';
    quote the averaged ri is '7' ;
proc/;
```

### Script A5. BARSG.run:
The script for calculating the biased average value of *r* in the FP analyses in Comparison 2 (see the main text)

```
macro= ;
macfloat 6;
    set 3 0;
    set 4 0;
loop 0 nchar
    if (!isact[#1]) continue; end
    set 1 length[0 #1]/weight[#1];
```

```
            set 9 maxsteps[#1];
            set 10 minsteps[#1];
            set 2 '9'-'1';
            set 7 '9'-'10';
             if 
((#1==0)||(#1==1)||(#1==2)||(#1==3)||(#1==28)||(#1==138)||(#1==7)||(
#1==8)||(#1==9)||(#1==10)||(#1==29)||(#1==141))
            set 5 1/6;
            set 6 '2'*'5';
            set 3 '3'+('5'*'7');
            set 4 '4'+'6';
            continue;
            end
            if ((#1==4)||(#1==5)||(#1==6)||(#1==139))
            set 5 1/4;
            set 6 '2'*'5';
            set 3 '3'+('5'*'7');
            set 4 '4'+'6';
            continue;
              end
               if 
((#1==11)||(#1==12)||(#1==13)||(#1==14)||(#1==15)||(#1==16)||(#1==30
)||(#1==142))
            set 5 1/8;
           set 6 '2'*'5';
             set 3 '3'+('5'*'7');
            set 4 '4'+'6';
            continue;
              end
              if 
((#1==17)||(#1==18)||(#1==20)||(#1==31)||(#1==34)||(#1==41)||(#1==35
)||(#1==36)||(#1==89)||(#1==39)||(#1==40)||(#1==117)||(#1==56)||(#1=
=61)||(#1==114)||(#1==67)||(#1==68)||(#1==69))
            set 5 1/3;
            set 6 '2'*'5';
            set 3 '3'+('5'*'7');
            set 4 '4'+'6';
            continue;
              end
              if 
((#1==19)||(#1==22)||(#1==21)||(#1==26)||(#1==37)||(#1==43)||(#1==48
)||(#1==49)||(#1==85)||(#1==87)||(#1==65)||(#1==66)||(#1==75)||(#1==
76)||(#1==77)||(#1==78))
            set 5 1/2;
           set 6 '2'*'5';
             set 3 '3'+('5'*'7');
             set 4 '4'+'6';
            continue;
               end
```

```
        if 
((#1==55)||(#1==64)||(#1==79)||(#1==82)||(#1==83)||(#1==84)||(#1==86
))
     set 5 1/7;
     set 6 '2'*'5';
     set 3 '3'+('5'*'7');
     set 4 '4'+'6';
     continue;
     end
    set 3 '3'+'7';
    set 4 '4'+'2';
    stop

   set 8 '4'/'3';
   quote the tilted averaged ri is '8' ;
proc/;
```

### Script A6. Onlyact
The scripts for activating only one out of the nine character groups, in the separate analyses of Comparison 1 (see the main text).

### Onlyact1.run

```
macro= ;
loop 0 nchar
    if (( maxsteps[#1] - minsteps[#1] )!=1)
     ccode ] #1;
    continue;
end
stop;
end ;
```

Note: for onlyact2.run, on Line 3 of the script, just replace the "1" at the right of the equation with "2", and so on.

### Script A7. Upact
The scripts for the serial analyses of "up-act" in Comparison 2 (see the main text)

### Upact1.run

```
macro= ;
loop 0 nchar
    if (( maxsteps[#1] - minsteps[#1] )<=1)
     ccode ] #1;
    continue;
end
stop;
end ;
```

Note: for upact2.run, on Line 3 of the script, just replace the "1" at the right of the inequality with "2", and so on.

**Script A8. Downact**
The scripts for the serial analyses of "down-act" in Comparison 2 (see text for explanations)

**Downact9.run**

```
macro= ;
loop 0 nchar
    if (( maxsteps[#1] - minsteps[#1] )>=9)
     ccode ] #1;
    continue;
end
stop;
end ;
```

Note: for downact8.run, on Line 3 of the script, just replace the "9" at the right of the inequality with "8", and so on.

**Part II. Notes for the data matrix**

**Data A1. SG04.tnt:**
The hominoid data matrix of the empirical example follows the source *(37)* directly. The 198 characters are composed of, in order, 89 "craniometric" characters *(37: 408: Table 3)* followed by 109 "traditional characters" *(37: 447: Appendix Table A1)*, with both following the sequence in the source *(37)*. The tnt data matrix is presented in the plain text format. Note that this is the matrix for the FP analysis, in which except for the downweighting adjustment for the redundant characters *(37: 426: Table 6)*, all steps receive equal weight. The step weight justification for the CP analyses needs to be achieved by using the script "CPSG.run".

```
nstates 16 ;
xread 'Data saved from TNT'
198 7
Colobus
2213244254515371712653624111285641635006853352112347777663440324024266323571545461533050220000030000010000?21220001200100000210010000010211000000100?000100011000100001221210100001011100000000100000
Papio
5545536362737583852823723325486500377317?6132576576385582312442617?210563724111170334113310000000300000000000?200200010113000020000130000200100000000100?000001001000100000122001000010111000000000130000
Hylobates
01014250413023301322735130022527427024?241060000202203646266130073322505243551345070647142000000300000010000?3122000120010010000001000001021200000000000001000010110010013110011000010011110100000300040
```

Pongo
5423442555333331313545424344583030054262816115536607604015285157545414562255426564234602230000010000201000030000000001000000000000000102000021000100000000000021101000020011101013001111110201001321

Gorilla
3314343543323350502714500132263023466444423276446626425400012074420200531442317442310512001000000000100000200000000001102002000010001010200003200000000000010000201101102022011010210000202110101111111

Pan
4435231422212220213432212133332433154464441434324424213453523346220202341344012453432221100000000000100000210000010001000000000001000000000000000000000000000010211101102020021211130000211121101012120

H.sapiens
1000000000000000010080002100010205510420000250010100061776706432707535200006400216656874520000020000240031101320002200313310222031110232002220010121131111011112211011020000202122311002010211000131300
;

Ccode
   +[/167 0       +[/167 1       +[/167 2       +[/167 3       +[/250 4
   +[/250 5       +[/250 6       +[/167 7       +[/167 8       +[/167 9
   +[/167 10      +[/125 11      +[/125 12      +[/125 13      +[/125 14
   +[/125 15      +[/125 16      +[/333 17      +[/333 18      +[/500 19
   +[/333 20      +[/500 21      +[/500 22      +[/1000 23     +[/1000 24
   +[/1000 25     +[/500 26      +[/1000 27     +[/167 28      +[/167 29
   +[/125 30      +[/333 31      +[/1000 32     +[/1000 33     +[/333 34
   +[/333 35      +[/333 36      +[/500 37      +[/1000 38     +[/333 39
   +[/333 40      +[/333 41      +[/1000 42     +[/500 43      +[/1000 44
   +[/1000 45     +[/1000 46     +[/1000 47     +[/500 48      +[/500 49
   +[/1000 50     +[/1000 51     +[/1000 52     +[/1000 53     +[/1000 54
   +[/143 55      +[/333 56      +[/1000 57     +[/1000 58     +[/1000 59
   +[/1000 60     +[/333 61      +[/1000 62     +[/1000 63     +[/143 64
   +[/500 65      +[/500 66      +[/333 67      +[/333 68      +[/333 69
   +[/1000 70     +[/1000 71     +[/1000 72     +[/1000 73     +[/1000 74
   +[/500 75      +[/500 76      +[/500 77      +[/500 78      +[/143 79
   +[/1000 80     +[/1000 81     +[/143 82      +[/143 83      +[/143 84
   +[/500 85      +[/143 86      +[/500 87      +[/1000 88     -[/333 89
   +]/1  90       +]/1  91       +]/1  92       +]/1  93       +]/1  94
   -[/1000 95     +]/1  96       +]/1  97       +]/1  98       +]/1  99
   +[/1000 100    +[/1000 101    +]/1  102      +]/1  103      +]/1  104
   +]/1  105      +]/1  106      +[/1000 107    +[/1000 108    +[/1000 109
   +[/1000 110    +]/1  111      +]/1  112      +]/1  113      +[/333 114
   +[/1000 115    +]/1  116      +[/333 117     +[/1000 118    +]/1  119
   +[/1000 120    +[/1000 121    -]/1  122      +[/1000 123    +[/1000 124
   +[/1000 125    +]/1  126      +]/1  127      +]/1  128      +[/1000 129
   +]/1  130      +]/1  131      +]/1  132      +]/1  133      -]/1  134
   +]/1  135      +[/1000 136    +[/1000 137    +[/167 138     +]/1  139
   +]/1  140      +[/167 141     +[/125 142     +]/1  143      +]/1  144
   +[/1000 145    +]/1  146      +[/1000 147    -]/1  148      +]/1  149
   +]/1  150      +]/1  151      +[/1000 152    +]/1  153      +[/1000 154
   +]/1  155      +[/1000 156    +[/1000 157    +]/1  158      +[/1000 159
   +[/1000 160    +]/1  161      +[/1000 162    +[/1000 163    +[/1000 164
   +[/1000 165    +[/1000 166    +[/1000 167    +[/1000 168    +[/1000 169
   +[/1000 170    +[/1000 171    +[/1000 172    +]/1  173      +[/1000 174
   +[/1000 175    +[/1000 176    +[/1000 177    +[/1000 178    +]/1  179
   +[/1000 180    +[/1000 181    +[/1000 182    +[/1000 183    +[/1000 184
   +[/1000 185    +[/1000 186    +[/1000 187    -[/1000 188    +[/1000 189
   +]/1  190      +[/1000 191    +[/1000 192    +[/1000 193    +[/1000 194
   +[/1000 195    -[/1000 196    +[/1000 197   ;

Ancstates
   -0        -1        -2        -3        -4        -5        -6        -7        -8        -9
   -10       -11       -12       -13       -14       -15       -16       -17       -18       -19

```
 -20      -21      -22      -23      -24      -25      -26      -27      -28      -29
 -30      -31      -32      -33      -34      -35      -36      -37      -38      -39
 -40      -41      -42      -43      -44      -45      -46      -47      -48      -49
 -50      -51      -52      -53      -54      -55      -56      -57      -58      -59
 -60      -61      -62      -63      -64      -65      -66      -67      -68      -69
 -70      -71      -72      -73      -74      -75      -76      -77      -78      -79
 -80      -81      -82      -83      -84      -85      -86      -87      -88      -89
 -90      -91      -92      -93      -94      -95      -96      -97      -98      -99
-100     -101     -102     -103     -104     -105     -106     -107     -108     -109
-110     -111     -112     -113     -114     -115     -116     -117     -118     -119
-120     -121     -122     -123     -124     -125     -126     -127     -128     -129
-130     -131     -132     -133     -134     -135     -136     -137     -138     -139
-140     -141     -142     -143     -144     -145     -146     -147     -148     -149
-150     -151     -152     -153     -154     -155     -156     -157     -158     -159
-160     -161     -162     -163     -164     -165     -166     -167     -168     -169
-170     -171     -172     -173     -174     -175     -176     -177     -178     -179
-180     -181     -182     -183     -184     -185     -186     -187     -188     -189
-190     -191     -192     -193     -194     -195     -196     -197     ;

xgroup

;
agroup

;
taxcode
+0       +1       +2       +3       +4       +5       +6
;

blocks 0;
tshrink
;
force

+
[ 2 3 4 5 6 ];

constrain=
;

proc/;
```

**Table A1. Details for each character group in the SG hominoid data matrix**.

| Character # | Effective count | Hp | Group # |
|---|---|---|---|
| 13 | 1/8 | 1 | 1 |
| 18 | 1/3 | 1 | 1 |
| 65 | 1/2 | 1 | 1 |
| 77 | 1/2 | 1 | 1 |
| 95 | 1 | 1 | 1 |
| 101 | 1 | 1 | 1 |
| 107 | 1 | 1 | 1 |
| 114 | 1/3 | 1 | 1 |
| 121 | 1 | 1 | 1 |
| 125 | 1 | 1 | 1 |
| 129 | 1 | 1 | 1 |
| 142 | 1/8 | 1 | 1 |
| 147 | 1 | 1 | 1 |
| 154 | 1 | 1 | 1 |
| 156 | 1 | 1 | 1 |
| 157 | 1 | 1 | 1 |
| 163 | 1 | 1 | 1 |
| 166 | 1 | 1 | 1 |
| 171 | 1 | 1 | 1 |
| 176 | 1 | 1 | 1 |
| 185 | 1 | 1 | 1 |
| 188 | 1 | 1 | 1 |
| 191 | 1 | 1 | 1 |
| 196 | 1 | 1 | 1 |
| 197 | 1 | 1 | 1 |
| 15 | 1/8 | 2 | 2 |
| 17 | 1/3 | 2 | 2 |
| 21 | 1/2 | 2 | 2 |
| 25 | 1 | 2 | 2 |
| 39 | 1/3 | 2 | 2 |
| 43 | 1/2 | 2 | 2 |
| 67 | 1/3 | 2 | 2 |
| 72 | 1 | 2 | 2 |
| 79 | 1/7 | 2 | 2 |
| 87 | 1/2 | 2 | 2 |
| 89 | 1/3 | 2 | 2 |
| 100 | 1 | 2 | 2 |
| 108 | 1 | 2 | 2 |

| Character # | Effective count | Hp | Group # |
|---|---|---|---|
| 117 | 1/3 | 2 | 2 |
| 118 | 1 | 2 | 2 |
| 120 | 1 | 2 | 2 |
| 123 | 1 | 2 | 2 |
| 124 | 1 | 2 | 2 |
| 137 | 1 | 2 | 2 |
| 141 | 1/6 | 2 | 2 |
| 145 | 1 | 2 | 2 |
| 152 | 1 | 2 | 2 |
| 160 | 1 | 2 | 2 |
| 162 | 1 | 2 | 2 |
| 164 | 1 | 2 | 2 |
| 165 | 1 | 2 | 2 |
| 167 | 1 | 2 | 2 |
| 172 | 1 | 2 | 2 |
| 174 | 1 | 2 | 2 |
| 175 | 1 | 2 | 2 |
| 180 | 1 | 2 | 2 |
| 181 | 1 | 2 | 2 |
| 182 | 1 | 2 | 2 |
| 184 | 1 | 2 | 2 |
| 186 | 1 | 2 | 2 |
| 189 | 1 | 2 | 2 |
| 192 | 1 | 2 | 2 |
| 193 | 1 | 2 | 2 |
| 195 | 1 | 2 | 2 |
| 5 | 1/4 | 3 | 3 |
| 10 | 1/6 | 3 | 3 |
| 23 | 1 | 3 | 3 |
| 24 | 1 | 3 | 3 |
| 28 | 1/6 | 3 | 3 |
| 37 | 1/2 | 3 | 3 |
| 44 | 1 | 3 | 3 |
| 50 | 1 | 3 | 3 |
| 61 | 1/3 | 3 | 3 |
| 64 | 1/7 | 3 | 3 |
| 70 | 1 | 3 | 3 |
| 73 | 1 | 3 | 3 |
| 138 | 1/6 | 3 | 3 |
| 159 | 1 | 3 | 3 |

| Character # | Effective count | Hp | Group # |
|---|---|---|---|
| 168 | 1 | 3 | 3 |
| 169 | 1 | 3 | 3 |
| 177 | 1 | 3 | 3 |
| 183 | 1 | 3 | 3 |
| 187 | 1 | 3 | 3 |
| 2 | 1/6 | 4 | 4 |
| 4 | 1/4 | 4 | 4 |
| 8 | 1/6 | 4 | 4 |
| 9 | 1/6 | 4 | 4 |
| 11 | 1/8 | 4 | 4 |
| 12 | 1/8 | 4 | 4 |
| 27 | 1 | 4 | 4 |
| 30 | 1/8 | 4 | 4 |
| 42 | 1 | 4 | 4 |
| 55 | 1/7 | 4 | 4 |
| 80 | 1 | 4 | 4 |
| 81 | 1 | 4 | 4 |
| 82 | 1/7 | 4 | 4 |
| 83 | 1/7 | 4 | 4 |
| 109 | 1 | 4 | 4 |
| 110 | 1 | 4 | 4 |
| 115 | 1 | 4 | 4 |
| 136 | 1 | 4 | 4 |
| 170 | 1 | 4 | 4 |
| 178 | 1 | 4 | 4 |
| 1 | 1/6 | 5 | 5 |
| 3 | 1/6 | 5 | 5 |
| 22 | 1/2 | 5 | 5 |
| 26 | 1/2 | 5 | 5 |
| 32 | 1 | 5 | 5 |
| 33 | 1 | 5 | 5 |
| 36 | 1/3 | 5 | 5 |
| 47 | 1 | 5 | 5 |
| 57 | 1 | 5 | 5 |
| 69 | 1/3 | 5 | 5 |
| 74 | 1 | 5 | 5 |
| 76 | 1/2 | 5 | 5 |
| 88 | 1 | 5 | 5 |
| 194 | 1 | 5 | 5 |
| 0 | 1/6 | 6 | 6 |

| Character # | Effective count | Hp | Group # |
|---|---|---|---|
| 6 | 1/4 | 6 | 6 |
| 41 | 1/3 | 6 | 6 |
| 54 | 1 | 6 | 6 |
| 60 | 1 | 6 | 6 |
| 62 | 1 | 6 | 6 |
| 71 | 1 | 6 | 6 |
| 85 | 1/2 | 6 | 6 |
| 7 | 1/6 | 7 | 7 |
| 14 | 1/8 | 7 | 7 |
| 19 | 1/2 | 7 | 7 |
| 20 | 1/3 | 7 | 7 |
| 34 | 1/3 | 7 | 7 |
| 35 | 1/3 | 7 | 7 |
| 58 | 1 | 7 | 7 |
| 59 | 1 | 7 | 7 |
| 63 | 1 | 7 | 7 |
| 68 | 1/3 | 7 | 7 |
| 75 | 1/2 | 7 | 7 |
| 84 | 1/7 | 7 | 7 |
| 29 | 1/6 | 8 | 8 |
| 40 | 1/3 | 8 | 8 |
| 46 | 1 | 8 | 8 |
| 48 | 1/2 | 8 | 8 |
| 49 | 1/2 | 8 | 8 |
| 51 | 1 | 8 | 8 |
| 53 | 1 | 8 | 8 |
| 56 | 1/3 | 8 | 8 |
| 66 | 1/2 | 8 | 8 |
| 78 | 1/2 | 8 | 8 |
| 16 | 1/8 | 9 | 9 |
| 31 | 1/3 | 9 | 9 |
| 38 | 1 | 9 | 9 |
| 45 | 1 | 9 | 9 |
| 52 | 1 | 9 | 9 |
| 86 | 1/7 | 9 | 9 |